\documentstyle[12pt]{article}
\newcommand{\bib}{\bibitem}

\newcommand{\spz}{\hspace{0.7cm}}

\newcommand{\nn}{\nonumber} 
\newcommand{\fr}{\rightarrow}
\newcommand{\de}{\partial} 
\newcommand{\ri}{\right} 
\newcommand{\lf}{\left}

\newcommand{\eq}{\begin{equation}}
\newcommand{\en}{\end{equation}}
\newcommand{\bea}{\begin{eqnarray}}
\newcommand{\eea}{\end{eqnarray}}

\newcommand{\ba}{\begin{array}}
\newcommand{\ea}{\end{array}}

\newcommand{\virg}{\spz,}
\newcommand{\pu}{\spz.}
\newcommand{\hg}{\hat{g}}
\newcommand{\hl}{\hat{l}}
\newcommand{\half}{\frac{1}{2}}

\newcommand{\M}{{\hbox{\large $\cal M \;$}}}

\newcommand{\T}{{\hbox{$Tr\;$}}}
\newcommand{\G}{{\hbox{\large $\cal G \;$}}}

\newcommand{\cH}{{\hbox{\large $\cal H \;$}}}
\newcommand{\AP}[1]{Ann.\ Phys.\ {\bf #1}}
\newcommand{\CMP}[1]{Comm.\ Math.\ Phys.\ {\bf#1}}

\newcommand{\JMP}[1]{Jou.\ Math.\ Phys.\ {\bf#1}}

\newcommand{\PR}[1]{Phys.\ Rev.\ {\bf #1}}
\newcommand{\PRL}[1]{Phys.\ Rev.\ Lett.\ {\bf #1}}

\begin{document}
\setlength{\unitlength}{.8mm}

\hfill  DFUL-1/02/96\\ \vskip 1cm
 \begin{center}  {\large \bf
Chern-Simons Field Theory and  Completely Integrable Systems } \\[.5cm]
\vskip 1.2cm 
 {\large  \it
L.Martina \footnote{E-mail: martina@le.infn.it}
	     , O.K.Pashaev$^{\dagger \,*}$ \footnote{E-mail:
pashaev@main1.jinr.dubna.su} and G.Soliani \footnote{E-mail:
soliani@le.infn.it}}\\
     
\vskip .8cm
  {\it Dipartimento di Fisica dell'Universit\'a and INFN
Sezione di 
 Lecce}\\
 {\it 73 100 Lecce, Italy}\\ $\dagger)$ {\it Joint Institute for
Nuclear Research, 141980 Dubna, 
 Russia}  \footnote{ Permanent address} \vskip 1cm 
\noindent {\bf Abstract} \\

\end{center} 
{\it  We show that  the classical non-abelian pure 
Chern-Simons  action is  related   in a natural way to   completely 
integrable systems of the Davey-Stewartson hyerarchy, via 
reductions of the gauge connection in  Hermitian
spaces and by performing certain gauge choices.  The B\"acklund 
Transformations are interpreted in terms of Chern-Simons  equations 
of motion or, 
on the other hand, as a consistency condition on the gauge. A mapping 
with a 
nonlinear $\sigma$-model is discussed.}
\vskip 1cm
\newpage

In the last ten years a great effort has been devoted in
the study of certain classes of nonlinear fields, namely
the Chern-Simons (CS) ~\cite{w1,w2}, the nonlinear
$\sigma$-model and  certain completely integrable models,
either at the quantum and in the classical level. There
exists a very wide literature concerning these topics and
several connections between them have been pointed out
from long time \cite{witten}. However, only recently a precise
relationship has been established at the classical level \cite{Comm}.
\par In this letter we show  that the (apparently
trivial) classical non-abelian CS field theory provides 
nonlinear
$\sigma$-models in a algorithmic way, including the basic 
structures  for their  integrability. In other words, we
provide a recipe for supplying completely integrable
models from  solvable models and what is the interplay
between the two concepts. 
\par

In doing so, for simplicity we restrict ourselves to  consider
the
$SU(2)$-CS theory, given by the action
 \eq
 S[J] = {k \over {4\pi}}\int_{\M} \T (J\wedge dJ +
\frac{2}{3} J\wedge J\wedge J), \label{1}  \en   where  $J$ is  the 
1-form 
 gauge connection with values in the  Lie  algebra 
$su\lf(2\ri)$  and ${\M}$ is a 3-manifold.  The classical equations of
motion for the action (\ref{1})  are given by  the  zero-curvature
condition  \eq
 F \equiv  dJ + J\wedge J = 0 \virg \label{eqmot}
 \en whose solutions are easily found in terms of chiral currents in 
the Lie group $SU\lf(2\ri)$.
 The action  (\ref{1})  is
 invariant under general coordinate transformations (preserving
orientation and volumes). Moreover, 
 under a  generic gauge map $G: \M  \fr SU\lf(2\ri)$ the gauge
connection transforms as usual by
 $ J \fr G^{-1}JG + G^{-1} dG $ . Correspondingly,  the  action
(\ref{1})  changes as
$ S[J] \fr S[J] +  2 \pi\, k\, W(G) , $
 where $W(G)$
 is the winding number of the map $G$ taking  integer values, as
prescribed by   the  homotopy theory 
\cite{Dub}. \par  Now, locally we trivialize ${\M}$  in the form
$\Sigma  \times
{\bf R}$, where
$\Sigma$ is a Riemann surface  and     $ {\bf R}$   is interpreted 
 as the time. This operation  breaks the general covariance of
the theory.
\par On the other hand, we can introduce a ${\bf Z}_{2}$ -  
decomposition
in the algebra of the 1-forms $J$.  In terms of the Lie algebra 
$\hg$ of
\G $(\,\equiv  SU\lf(2\ri)$  in the  case we are considering), this   
means that  
 $
\hg = \hl^{(0)} \oplus
\hl^{(1)}, \;
 [ \hl^{(i)}, \hl^{(j)}] \subset  \hl^{(i+j)\; {\hbox{mod (2)}}  }$,
 where
$\hl^{(0)}$  is the Lie algebra of a proper subgroup \cH  of \G and
$\hl^{(1)}$  is the 
  complement of  $\hl^{(0)}$ in $\hg$. The subgroup \cH  is chosen to
 be
invariant under an involution over \G.  The group \G acts transitively
on the coset space
\G/\cH, which   is  a symmetric space \cite{Hel}. 
    At any  point
$p_{0}   \in$ 
\G/\cH,  the  tangent
 space $T_{p_{0}}\left({\G/\cH }\ri)$   is isomorphic to   $\hl^{(1)}$.
The natural Riemann connection defined on these spaces is torsionless.
Moreover, we require the existence of a complex structure on such a
space,  thus we  deal with a    Hermitian symmetric space
\cite{Hel}. \par  The current $J$ is decomposed   in  the form
\eq  J = J^{(0)}+ J^{(1)} \virg                                        
\label{dec}
\en where $J^{(1)} \;$ and  $J^{(0)} \;$ are 1-forms  taking  values in
the tangent space and in the isotropy algebra of the considered 
hermitian
space, respectively. The two components of the current obtained by the
previous decomposition will play different roles in the theory. 
Indeed, we
will see that the model  (\ref{1}) will become a non-relativistic 
theory for some matter fields minimally coupled to a residual CS
gauge field (in
the sense that it is associated with  the isotropy group \cH ), 
plus some
 contraints expressing the torsionless
character of the target space \G/\cH. Indeed, if for \G$ \equiv
SU\lf(2\ri)$ we 
 choose \cH  $\equiv \,U\lf(1\ri)$, the related  hermitian space is
the sphere $ SU\lf(2\ri)/U\lf(1\ri) \cong {\cal S}^{2} 
\cong CP^{1} $ and the action (\ref{1}) can be rewritten in the form
\bea   S  =   -  \frac{k}{ \pi} \int_{\Sigma  \times {\bf R}}  
\left\{{    \half  \epsilon^{\lambda\, \mu\,\nu} \, v_{\lambda}
\de_{\mu}\,v_{\nu} } \right.\hskip 3cm \nn
 \label{2}  \\
     \left.{ +   i \half  \lf(\psi ^{*}_{+}  {  D}_{0}\psi _{+} - \psi
_{+}  \lf({  D}_{0}\psi _{+}\ri)^{*}
  - \psi ^{*}_{-}   {  D}_{0}\psi _{-} + \psi _{-} \lf({  D}_{0}\psi
_{-}\ri)^{*}\ri)  } \right. \hskip 1cm  \\
\left.{ - i q^{*}_{0} \lf({   D}\psi _{+} - {\bar{  D}}\psi _{-} \ri) 
+ 
i q_{0} \lf({ D}\psi _{+}  - {\bar{ D}}\psi _{-} \ri)^{*}  }\right\}
 \,dx^{0}\,dx^{1}\,dx^{2}  \nn
\virg
\eea where the  fields $\psi_{\pm}$ parametrize the space components of
$J^{(1)}$ (they  can be  considered as scalar complex matter fields),
 $v_{\mu}$ parametrizes $J^{(0)}$ and represents the  abelian CS gauge
field associated with the $U\lf(1\ri)$ invariance. Finally, the field
$q_{0}$, related to the time component of $J^{(1)}$,  plays
 the role of a  Lagrange multiplier, enforcing a constraint, which is
the remnant of the torsionless property of the chosen target space.
$D_{0} = \de_{0} - 2 i v_{0}$, $ D = \de_{z}  - 2 i v $ and
$\bar{ D} = \de_{\bar{z}}  - 2 i v^{*} $  denote  covariant 
derivatives,
 where we have employed the usual complex
variables $z = x_{1} + i x_{2}$ and $\bar{z} = x_{1} - i x_{2}$ and  
with
 $v = \half \lf( v_{1} - i v_{2}\ri)$.
 By resorting to higher dimensional
compact group \G, we get analogous structures, in which we can embed
several types of non-relativistic $CP^{n}$  models.  Moreover, for the 
system described by (\ref{2}), we studied the canonical structure,
which turns out to be a completely constrained Hamiltonian system
\cite{Comm}. \par However, the previous approach contains also an
unexpected structure. In fact, among the variations of  the action 
(\ref{2})
 let us take for brevity only the equations
\bea
{\bar D}{\psi }_{-}=D{\psi }_{+}
\label{3a} \virg  \\ 
{\partial }_{z}{v}^{*}-{\partial
}_{\overline{z}}v=-i\left({{\left|{{\psi
}_{+}}\right|}^{2}-{\left|{{\psi }_{-}}\right|}^{2}}\right) \virg
\label{3b} 
\eea
which are the torsionless condition and the
Gauss-Chern-Simons law (GCS), respectively. These
equations are very important, because they are the unique  equations 
which 
do not contain
time derivatives and time-components of the currents
$J$. 
\par We handle Eqs. (\ref{3a} - \ref{3b}) by the help of
the   new matrix fields
\bea
\matrix{\mit {\cal  V} \rm =\left({\matrix{{v}^{*}&\cr
&v\cr}}\right)&{\hat{\Psi }}_{\pm }=\left({\matrix{&{\psi }_{\pm }\cr
-{\psi }_{\pm }^{*}&\cr}}\right)\cr}\pu \label{4}
\eea
Furthermore, let us introduce 
\eq
{\rm B}^{\left({1}\right)}={i \over 2}{\sigma }_{3}\left({{\hat{\Psi
}}_{-}-{\hat{\Psi }}_{+}}\right) \pu
\en
Combining the GCS law in Eq. (\ref{3b}) with its complex
conjugate, we obtain
\eq
\rm Tr\left\{{{\sigma }_{3}\left[{\left({\matrix{{\partial }_{z}&\cr
&{\partial }_{\overline{z}}\cr}}\right) {\cal  V} +
{B}^{\left({1}\right)}
{\hat{\Psi }}_{-}-{\hat{\Psi
}}_{+}{B}^{\left({1}\right)}}\right]}\right\}=0 \pu
\en
 Since the quantity in the square brackets is a diagonal matrix and no
 information is supplied about the identity component, we have the
relation 
\eq
{\left({\matrix{{\partial }_{z}&\cr
&{\partial }_{\overline{z}}\cr}}\right) {\cal  V} +
{B}^{\left({1}\right)}
{\hat{\Psi }}_{-}-{\hat{\Psi
}}_{+}{B}^{\left({1}\right)}} = f \sigma_{0} \virg \label{5}
\en
where $f = f\lf(z, \bar{z}\ri)$ is an arbitrary  function and
$\sigma_{0}$ is the identity matrix.\par On the other hand, the
torsionless condition can be written as

\eq
{\left({\matrix{{\rm \partial }_{z}&\cr
&{\partial }_{\overline{z}}\cr}}\right){B}^{\left({1}\right)}+
{i \over 2}\left({{\partial }_{\overline{z}}-{\partial
}_{z}}\right){\hat{\Psi }}_{-}+{\cal V} {\hat{\Psi }}_{-} - {\hat{\Psi
}}_{+}{\cal V}  =0} \pu \label{6}
\en
Equations (\ref{5}) and (\ref{6}) can be summed up  to give the  expression
\eq 
{\rm T}_{+}\left[{{i \over 2}\left({{\partial
}_{\overline{z}}-{\partial
}_{z}}\right)+{\cal V}+{B}^{\left({1}\right)}}\right]-\left[{{i \over
2}\left({{\partial }_{\overline{z}}-{\partial
}_{z}}\right)+{\cal V}+{B}^{\left({1}\right)}}\right]{T}_{-}=f ,
\label{7} \en
where 
\eq {\rm T}_{\pm }=\left({\matrix{{\partial }_{z}&\cr
&{\partial }_{\overline{z}}\cr}}\right) -{\hat{\Psi }}_{\pm } \pu
\label{8} \en 
Since the previous procedure is invertible (the summation is made over
  independent
components), Eq. (\ref{7}) is equivalent to the system
(\ref{3a} - \ref{3b})  (modulo $f$).\par
Putting  $f  \equiv 0$,  Eq.  (\ref{7}) coincides with the
space  part of the B\"acklund transformation for the two-dimensional
 Zachkarov-Shabat
problem \cite{Konob},  in which the principal spectral problem is 
given by 
 Eq. (\ref{8}), and the first order B\"acklund-gauge
operator is 
\eq
B = \nabla+{\cal V}+{B}^{\left({1}\right)}\virg \label{B-g}
\en  
with  $\nabla = 
{{i \over 2}\left({{\partial
}_{\overline{z}}-{\partial }_{z}}\right)}$. The operator
$B$ transforms an eigenfunction
$\phi_{-}$ of the linear problem $T_{-} \phi_{-} =0$ into
$\phi_{+} = B  \phi_{-}$, which is an eigenfunction of
$T_{+} \phi_{+} =0$.

It is well known \cite{Konob} that the triad $B$, $T_{\pm}$
enables  one to introduce a continuous extra-dependency on a
parameter, say $\tau$, and two operators of the form 
$T^{\left(\tau\ri)}_{\pm} =  i \partial_{\tau} + \sum_{k
=0}^{N} T_{\pm,\; k}
\nabla^{N-k}$  such
that:  1) $\lf[ T_{\pm} , \; T_{\pm}^{\left(\tau\ri)} \ri]
= 0$ and 2)  $T_{+}^{\left(\tau\ri)} B - B
T_{-}^{{\left(\tau\ri)}} = 0$. The lowest order operator
of such a type, leading to a non-trivial equation, can be
put in the form
\eq  T^{\left(\tau\ri)}_{\pm} =  i \partial_{\tau} -
8 \sigma_{3}
\nabla^{2} - 8 i{\hat \Psi}_{\pm}\nabla + 4
\left({\matrix{{\partial }_{\bar z}&\cr & -{\partial
}_{z}\cr}}\right) {\hat
\Psi}_{\pm} - 2 \eta_{\pm} \sigma_{0} + {1
\over 4} \zeta_{\pm}
\sigma_{3} \pu \label{tprob} 
\en
and 
 the corresponding evolution equation reads
\bea
 i \partial_{\tau} \psi_{\pm} + 2 \lf(\de_{z}^{2}
+\de_{\bar z}^{2}\ri)\psi_{\pm} + {1 \over 2} \zeta_{\pm}
\psi_{\pm} = 0 \virg \nn \\ \de_{z} \de_{\bar z}
\zeta_{\pm} =
 8 \lf(\de_{z}^{2} +\de_{\bar z}^{2}\ri) |\psi_{\pm}|^{2}
\virg
\\
\de_{z}
\de_{\bar z} \eta_{\pm} = \lf(\de_{z}^{2} -\de_{\bar
z}^{2}\ri) |\psi_{\pm}|^{2} \virg \nn \label{DS}
 \eea 
which is the Davey-Stewartson II equation (DS II)
\cite{Konob}. We summarize the first result saying that
when we embed  the classical ``trivial'' CS theory into a
special geometric setting (we chose the  special
trivialization $\Sigma \times {\bf R}$ for the space-time
and a  hermitian space as target space), some  topics
related to the completely integrable models appear.
However, at this stage the construction is not complete at
all. Indeed if we look at the evolution equations arising
from the action (\ref{2}), we have 
\bea
D_{0} \psi_{+} -
 {\bar D} q_{0}=0 \nn \virg \\
D_{0} \psi_{-} -
  D q_{0}=0 \pu \label{evoleq1}
\eea
In its turn the evolution of $q_{0}$ and of $v_{0}$ is completely 
arbitrary, since these quantities are Lagrange multipliers 
associated with
 the gauge degrees of freedom. 
This structure is quite different from the DS equation, unless
 we break the general gauge invariance of the CS theory, requiring 
a constraint
 on $q_{0}$. Which is the form of such a constraint? \par
The simplest choice (Weyl gauge) is provided by 
 $q_{0}=v_{0} \equiv 0$, which  has been widely exploited (\cite{w2},
 \cite{dun}). 
In this gauge the Lagrangian becomes quadratic and the quantization  
using 
the canonical formalism can be performed. However, here we want to
explore other gauge choices and their consequences at the classical
level.\par
For instance, let us put 
\eq
 {q}_{0}=2i\left[{\left({\overline{D}+
{i \over 2}{w}_{+}}\right){\psi }_{+} +
\left({D+{i \over 2}{w}_{-}}\right){\psi }_{-}}\right] 
\virg \label{cond1} 
\en
where we assume that the complex function $w_{+} =
w_{-}^{*}$ is invariant under and a $U\lf(1\ri)$ gauge
transformation acting on the CS - fields.  Moreover,  accordingly
to them   $\lf( w_{+}, w_{-}\ri)$ transforms controvariantly 
under space-time
transformations. The constraint (\ref{cond1}) admits 
$U(1)$ 
 as residual gauge symmetry. At the same time the general
covariance symmetry is broken and only special Lie-point
symmetries are allowed.
 A detailed analysis of this aspect is skipped for the
moment.
\par The substitution of Eq. (\ref{cond1}) into the CS-
field equations provides certain nonlinear evolution
equations for $\psi_{\pm}$. In particular,   Eqs.  (\ref{evoleq1}) 
become
\bea
D_{0}\psi_{+} - 2i \lf( D^{2} +{\bar D}^{2}\ri) \psi_{+}+ 
 \psi_{+} \de_{\bar z}w_{+}
+w_{+} {\bar D}\psi_{+} + w_{-} D\psi_{+} =\nn \\ 
\lf[ 4i \left({{\left|{{\psi
}_{+}}\right|}^{2}-{\left|{{\psi }_{-}}\right|}^{2}}\right) -
  \de_{\bar z}w_{-}\ri] \psi_{-}, \nn \\
D_{0}\psi_{-} - 2i \lf( D^{2} +{\bar D}^{2}\ri) \psi_{-}+  
\psi_{-} \de_{\bar z}w_{-}
+w_{+} {\bar D}\psi_{-} + w_{-} D\psi_{-} =\nn \\  - 
\lf[ 4i \left({{\left|{{\psi
}_{+}}\right|}^{2}-{\left|{{\psi }_{-}}\right|}^{2}}\right) +
  \de_{ z}w_{+}\ri] \psi_{+}. \label{preSch}
\eea
Moreover,  we have to take into account Eqs (\ref{3a} -
\ref{3b}) and  the equation involving the time-derivatives
of
$v,\; {\bar v}$ (the ``electric strenght'' field), which we
write for an  arbitrary $q_{0}$
\bea
\de_{0} v -\de_{z} v_{0} =i \lf( q_{0} \psi_{+}^{*} - 
q_{0}^{*} \psi_{-}\ri)\virg\nn \\
\de_{0} {\bar v} -\de_{\bar z} v_{0} = - i \lf( q_{0}^{*} \psi_{+} - 
q_{0} \psi_{-}^{*}\ri) \pu \label{electr}
\eea
We notice that in  Eqs. (\ref{preSch}) the coupling between
the  components  $\psi_{\pm}$  is nonlocal through the
$U\lf(1\ri)$ gauge fields,  and local by the r.h.s.. The quantity
$w_{\pm}$ is a sort of external  field. 
\par However, inspired by the previous discussion on the DS
equation, we  can switch  off the  local coupling  just by
putting
\eq
\de_{z} w_{+} = - 4i  \left({{\left|{{\psi
}_{+}}\right|}^{2}-{\left|{{\psi }_{-}}\right|}^{2}}\right), \label{w}
\en
which implies a sort of Gauss law like  Eq. (\ref{3b}) 
 containing also the  zero-divergence condition. Then, we
 keep 
$v$ and
$w_{-}$ still distinct. On the other hand, exploiting this 
similarity, we
can combine them into  the irrotational field 
${ A} =4v - {1\over 2} w_{-}^{T}$, with $w_{-} = w_{-}^{L}
 + w_{-}^{T} =
 -  \de_{z}\omega -  i
 \de_{z}\chi $. The stream funtion $\chi$ and the potential $\omega$
 satisfy  the Poisson and the Laplace  equations, respectively 
\bea 
\de_{z} \de_{\bar z} \chi = - 4 \left({{\left|{{\psi
}_{+}}\right|}^{2}-{\left|{{\psi }_{-}}\right|}^{2}}\right) 
\label{Poi}  \\
\de_{z} \de_{\bar z} \omega =0 \label{Lap}  \pu
\eea
Exploiting the $U\lf(1\ri)$ gauge invariane   of   Eqs.
(\ref{preSch}) and (\ref{electr}), we can make
the substitution 
\eq A = \de_{z} \Lambda \;\lf( \Lambda  
\in {\bf R}\ri),
\qquad \psi_{\pm} =  \Psi_{\pm} e^{{i\over 2}
\Lambda}, \qquad  v_{0} = {1 \over 4} \lf( A_{0} +
\de_{0} \Lambda\ri) \virg \label{lambda}
\en where $A_{0}$ is a new time component of the
CS
$U\lf(1\ri)$-scalar field. In this formalism the
 condition (\ref{3a}) reads
\eq
 \left({{\partial }_{\bar{z}}+{1 \over 4}{\partial
}_{\bar{z}}\chi }\right){\Psi }_{-}=\left({{\partial
}_{z}-{1 \over 4}{\partial }_{z}\chi }\right){\Psi }_{+} \label{tor}
\en
 and the "electric" field equations (\ref{electr})
become a pair of compatible first order equations for
$A_{0}$ in terms of $\Psi_{\pm}$, their derivatives and
derivatives of
$\chi$.
\par
However, it is convenient to introduce the quantities
\bea
{A}_{0}^{\lf(\pm \ri) }={A}_{0} \mp \left.{\left({{\partial
}_{z}^{2}\chi +{\partial }_{\bar{z}}^{2}\chi
}\right)-{1 \over 4}\left({{\left({{\partial }_{z}\chi
}\right)}^{2}+{\left({{\partial }_{\bar{z}}\chi
}\right)}^{2}}\right)}\right.+ \nn \\
{i \over
2}\left({{\partial }_{z}\omega {\partial }_{z}\chi
-{\partial }_{\bar{z}}\omega {\partial }_{\bar{z}}\chi
}\right)-2i\left({\matrix{{\partial
}_{\bar{z}}^{2}\cr {\partial
}_{z}^{2}\cr}}\right)\omega , \label{auxf} 
\eea
which allows us  to write the time evolution for $\Psi_{\pm}$
in the form 
\eq
i{\partial }_{0}{\Psi }_{\pm }+2\left({{\partial
}_{z}^{2}+{\partial }_{\bar{z}}^{2}}\right){\Psi
}_{\pm }+{1 \over 2}{A}_{0}^{\lf(\pm \ri) }{\Psi }_{\pm }
- i \lf( \de_{\bar z} \omega \de_{\bar z} + \de_{z}
\omega \de_{z} \ri) \Psi_{\pm} = 0\pu \label{genDS}
\en
 From Eq. (\ref{auxf}) we notice that the functions 
${A}_{0}^{\lf(\pm \ri) }$ are not independent, but they
are related by
\eq
A_{0}^{+} -  A_{0}^{-} = - 2\lf( \de_{\bar z}^{2} +
\de_{ z}^{2} \ri) \chi -2 \lf( \de_{\bar z}^{2} -
\de_{ z}^{2} \ri) \omega \pu \label{algconst}
\en
Moreover, they satisfy the  equation
\eq
{\partial }_{z}{\partial
}_{\overline{z}}{A}_{0}^{\left({\pm
}\right)}=8\left({{\partial }_{z}^{2}+{\partial
}_{\overline{z}}^{2}}\right){\left|{{\Psi }_{\pm
}}\right|}^{2}\virg  \label{auxeq}
\en
which is a  consequene of the electric
strenght (\ref{electr}).
\par
To summarize,  by the specific gauge
choices (\ref{cond1}) and (\ref{lambda}) we have obtained
a formally decoupled pair of DS - like   equations   for the
fields
 $\lf( \Psi_{\pm},{A}_{0}^{\lf(\pm \ri) }\ri)$. A
generalizing term, involving first order derivatives of
$\Psi_{\pm}$, has  coefficients depending on the
harmonic map $\omega$. For $\omega = const$ we recover
the system (16).
Actually, Eqs. (\ref{Poi}), (\ref{tor}) and 
(\ref{algconst}) close the system, introducing a non-local
coupling. But Eqs. (\ref{Poi}) and  (\ref{tor}) are essentially the 
system (\ref{3a} - \ref{3b}) discussed at the
beginning,  providing  the space part of the
B\"acklund transformations for the DS system.
In other words,  we have obtained  a pair of DSII systems
 coupled by the B\"acklund transformations. This result
can be  used  in looking for classes  of solutions
for the CS theory in the special gauge (\ref{cond1}), 
following the standard methods developed in the context
of the completely  integrable systems. For instance, for 
$\Psi_{+} \equiv 0$ we can find  $\Psi_{-}$ in terms of
solutions of the Liouville equation, to which our system
of equations reduces. Such  solutions of
multivortex type are  widely discussed in
\cite{CSF,Arkadiev}. 
\par
The  discussion above  can be extended to the generalized DSII
system (i.e. for non-constant harmonic background
$\omega$).     In fact the system given by Eqs.
(\ref{genDS} - \ref{auxeq}) (for instance,  let us
consider the "-" case) admits as  Lax pair  the operator
$T_{-}$ defined  in (\ref{8}) and the generalization of
(\ref{tprob})
\bea  T^{\left(\tau\ri)}_{-} =  i \partial_{0} -
8 \sigma_{3}
\nabla^{2} - 8 \lf( i{\hat \Psi}_{-} + R \ri) \nabla + 4
\left({\matrix{{\partial }_{\bar z}&\cr & -{\partial
}_{z}\cr}}\right) {\hat
\Psi}_{-} \nn \\
- 4i  \sigma_{3} \lf[R,{\hat
\Psi}_{-}\ri] - 2 \eta_{-}
\sigma_{0} + {1
\over 4} \lf( A_{0}^{\lf(- \ri)} - 2 i \de_{z}^{2}
\omega\ri) \sigma_{3} \virg
\eea
where    
\eq
R={1 \over 4}\left({\matrix{{\partial
}_{\overline{z}}\omega &\cr &{-\partial }_{z}\omega
\cr}}\right) \pu
\en
Furthermore, in analogy with the DS equation \cite{noi},
one can look for a  gauge transformation between  the
above system to a  spin model. In fact, introducing the spin field
 $S$ ($S \in  SU\lf(2\ri)/U\lf(1\ri)$) one can easily prove that the 
system 
(in real variables)
\bea
\de_{0} S + \Re\lf(w_{+}\ri) \de_{1}S - \Im\lf(w_{+}\ri) \de_{2}S
+{i \over 2} \lf[S, \lf(\de_{1}^{2}-\de_{2}^{2}\ri)S\ri] = 0 \virg 
\nn \\
\de_{1}  \Re\lf(w_{+}\ri) + \de_{2}  \Im\lf(w_{+}\ri) = 0 \virg  \\
\de_{1}  \Re\lf(w_{+}\ri) - \de_{2}  \Im\lf(w_{+}\ri) =
- {i \over 2} Tr\,{\lf( S \lf[\de_{1} S, \; \de_{2} S\ri]\ri)} \pu \nn
\label{TopmagI} \eea
is equivalent to (\ref{Poi}-\ref{Lap}-\ref{genDS}-\ref{auxeq})
 (only one pair of fields, for instance  $\lf( \Psi_{-}, 
\, A_{0}^{\lf(-\ri)}\ri)$,  is kept).
 This can be seen by looking for  a suitable non-degenerate matrix $g$,
 such that the Lax pair of the spin model
\bea
L=i{\partial }_{2}+S{\partial
}_{1} \virg \nn \\
 M = {\partial }_{0}+2 i S{\partial
}_{1}^{2}+ \lf( i \de_{1} S + S  \de_{2}S -  \Im\lf(w_{+}\ri)S + 
 \Re\lf(w_{+}\ri)\ri){\partial }_{1}
\eea
takes  the form 
\eq
{T}_{-}= {g}^{-1} L g, \qquad {T}_{-}^{\left({\tau}\right)} = g^{-1} 
M g,
 \qquad  S = g \sigma_{3} g^{-1}\pu
\en

Then we can interprete this system as a two dimensional continuous 
spin 
field in a moving frame, determined by the incompressible velocity 
field
 $\Re\lf(w_{+}\ri), \, \Im\lf(w_{+}\ri)$ in a non-euclidean space 
metric 
$(+, - )$. The vorticity is determined by the density of the 
topological 
charge. These types of  systems  have
 been  discussed in \cite{CSF} - 
\cite{bilin}. They  can be considered  as  generalizations
 of the 
well-known Ishimori model \cite{Ishi}.
In this context it is interesting to observe that the diagonal 
element of the B\"acklund operator (\ref{B-g}) takes  a physical 
meaning.  Finally, we  notice that the system (\ref{TopmagI}) 
can be treated by resorting to the tangent space representation 
approach \cite{CSF}. This method allows  to describe the spin model 
(\ref{TopmagI}) in terms of a non-relativistic gauge theory. In 
the specific  case one obtains the full system  ( \ref{Poi}- 
\ref{Lap} -\ref{genDS}- \ref{auxeq}). If we  consider a 
generalization of the above liquid spin model, by  
introducing an arbitrary non-vanishing coupling 
constant $\theta$ between the vorticity and the 
topological density in the third equation of 
(\ref{TopmagI}),  the resulting system is not 
longer integrable and can be analyzed only by the 
help of the tangent space representation approach. 
However, the main claim is that one cannot combine 
the velocity field and a suitable gauge field into 
into an irrotational field, like $A$. Such a ``phenomenological'' 
model  could be related to the creation of vortices in the 
superfluid $^{3}He$ \cite{Mermin}, but it  is not reducible 
in the framework of the $SU\lf(2\ri)$-CS theory developed in this work.
\par

This work was supported in part by MURST
of Italy and by INFN - Sezione di Lecce. One of the authors
 (O. K. P.) thanks
the Department of Physics of Lecce University for the warm hospitality.
\vfill
\newpage

\end{document}